\nonstopmode

\documentclass[prl,aps,twocolumn,showpacs,superscriptaddress]{revtex4}
\usepackage{bm,amsmath,amssymb,graphicx}

\begin{document}
\title{Quantum state-resolved probing of strong-field-ionized xenon atoms using
femtosecond high-order harmonic transient absorption spectroscopy}
\author{Zhi-Heng Loh}
\affiliation{Departments of Chemistry and Physics, University of California,
Berkeley, CA 94720, USA, and Chemical Sciences Division, Lawrence Berkeley
National Laboratory, Berkeley, CA 94720, USA}
\author{Munira Khalil}
\affiliation{Departments of Chemistry and Physics, University of California,
Berkeley, CA 94720, USA, and Chemical Sciences Division, Lawrence Berkeley
National Laboratory, Berkeley, CA 94720, USA}
\author{Raoul E. Correa}
\affiliation{Departments of Chemistry and Physics, University of California,
Berkeley, CA 94720, USA, and Chemical Sciences Division, Lawrence Berkeley
National Laboratory, Berkeley, CA 94720, USA}
\author{Robin Santra}
\affiliation{Argonne National Laboratory, Argonne, IL 60439, USA}
\author{Christian Buth}
\thanks{Self-employed, Germany}
\affiliation{Argonne National Laboratory, Argonne, IL 60439, USA}
\author{Stephen R. Leone}
\email{srl@berkeley.edu}
\affiliation{Departments of Chemistry and Physics, University of California,
Berkeley, CA 94720, USA, and Chemical Sciences
Division, Lawrence Berkeley National Laboratory, Berkeley, CA 94720, USA}
\date{12 April 2007}

\begin{abstract}
Femtosecond high-order harmonic transient absorption spectroscopy is used to
resolve the complete $\vert j,m \rangle$ quantum state distribution of Xe$^{+}$
produced by optical strong-field ionization of Xe atoms at $800$~nm. Probing at
the Xe $N_{4/5}$ edge yields a population distribution $\rho_{j,\vert m \vert}$
of $\rho_{3/2,1/2}:\rho_{1/2,1/2}:\rho_{3/2,3/2} = 75 \pm 6 : 12 \pm 3 : 13 \pm
6 \%$. The result is compared to a tunnel ionization calculation with the
inclusion of spin-orbit coupling, revealing nonadiabatic ionization behavior.
The sub-50-fs time resolution paves the way for table-top extreme ultraviolet
absorption probing of ultrafast dynamics.
\end{abstract}

\pacs{42.50.Hz, 32.80.Rm, 42.65.Ky, 82.53.-k}
\maketitle

Studies of laser-atom interactions in the nonperturbative, strong-field regime
elucidate novel phenomena such as above-threshold ionization
\cite{Agostini79,Corkum89}, nonsequential double ionization \cite{DiMauro94},
and high-order harmonic generation  \cite{McPherson87,Balcou93,Gordon93}. While
these processes are extensively studied both experimentally and theoretically,
details remain unknown about the $\vert j,m \rangle$ state distribution of the
photoion produced by the initial photoionization step ($m$ is the projection
quantum number associated with the total angular momentum $j$ of the hole).
Moreover, experimental tests of theoretical models for strong-field ionization
mostly rely on measuring the ion yield as a function of laser peak intensity
\cite{Chin91}. The strong dependence of ionization yields on the orbital angular
momentum and its direction relative to the laser polarization axis, as predicted
by most theoretical models (e.g., the Ammosov-Delone Krainov (ADK) rates
\cite{Ammosov86}), suggests that knowledge of the complete $\vert j,m \rangle$
state distribution can be used as an additional benchmark for theory. Young {\it
et al.} recently reported the use of synchrotron x-ray pulses to probe the
hole-orbital alignment of Kr$^{+}$ photoions generated in the strong-field
ionization of Kr \cite{YoAr06}. The unresolved fine-structure transitions
prevented retrieval of the complete $\vert j,m \rangle$ state distribution.
However, the observed degree of alignment is reproduced by the $\vert j,m
\rangle$ state distribution obtained by tunnel ionization calculations with the
inclusion of spin-orbit coupling \cite{SaDu06}.

Here we investigate the experimental and theoretical strong-field ionization of
xenon to extract the complete $\vert j,m \rangle$ quantum state distribution.
Femtosecond extreme ultraviolet (EUV) transient absorption spectroscopy is
demonstrated with a laser-based, high-order harmonic probe source for the
experiments; results are compared to tunnel ionization calculations that
incorporate spin-orbit coupling. The resultant angular momentum distribution and
hole-orbital alignment of the Xe$^{+}$ photoions are measured by probing the
transition from the $4d$ core level to the $5p$ valence shell. These
measurements allow the determination of the complete $\vert j,m \rangle$ quantum
state distribution, which is compared to theory.

\begin{figure}[b]
  \includegraphics[width=8cm,origin=c]{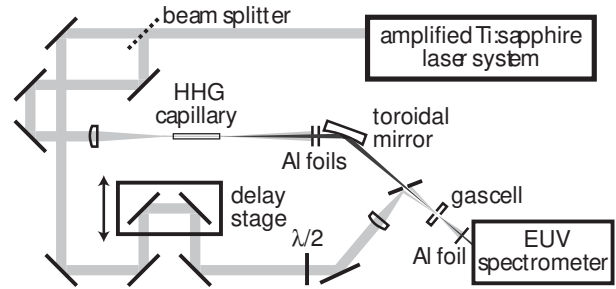}
  \caption[]{Schematic illustration of the experimental setup. The light and dark
           lines correspond to the 800-nm pump and EUV probe beam paths, respectively.}
  \label{fig1}
\end{figure}

The schematic of the experimental setup is illustrated in Fig.~\ref{fig1}.
Briefly, the amplified output from a commercial Ti:sapphire laser system
($2.4$~W, $800$~nm, $45$~fs, $1$~kHz) is sent to a $20:80$ beamsplitter to
produce the optical pump and high-order harmonic generation beam, respectively.
High-order harmonics in the EUV region are generated by focusing the laser light
into a $7$-cm-long, $150$-$\mu$m internal diameter capillary filled with $6.0
\times 10^{3}$ Pa of neon \cite{KaMu98}.  The estimated photon flux at the
source is $10^{4}$ photons per pulse for the high-order harmonic centered at
$55.4$~eV. A pair of 0.2-$\mu$m-thick Al foils is used to reject the residual
$800$-nm light and transmit the high-order harmonics. After reflection by a
toroidal mirror, the high-order harmonics are refocused into a 2-mm-long gas
cell filled with $2.7 \times 10^{3}$ Pa of Xe \cite{pressure}. Scanning
knife-edge measurements give a beam waist of 21~$\mu$m for the high-order
harmonics. The transmitted EUV radiation is spectrally dispersed in a home-built
spectrometer and detected with a thermoelectrically-cooled CCD camera. A
dielectric-coated mirror with a 1-mm-diameter internal bore hole allows the
optical pump beam to overlap with the EUV probe beam in a collinear geometry. A
half-waveplate inserted into the path of the pump beam allows the relative
polarization of the pump and probe beams to be varied. The polarization purity
of the pump beam is characterized by an extinction ratio of $>200:1$. The 800 nm
pump pulse energy incident on the gas cell is $0.13$~mJ. Further measurements
confirm that the pump pulse energy after the gas cell remains unchanged (by
$<3\%$) upon rotation of the half-waveplate. In the presence of the Xe gas, the
pump pulse duration increases slightly to $49$~fs and the elliptical focal spot
of the pump beam is characterized by horizontal and vertical beam waists of 61
and 34 $\mu$m, respectively; these parameters yield a peak intensity of $8\times
10^{13}$~W/cm$^2$ for the pump pulse. The pump-probe time delay is varied by
means of an optical delay line in the path of the pump beam. Transient
absorption spectra are obtained by using spectra collected at $-500$~fs time
delay as the reference \cite{pumpbeam}; a negative time delay implies that the
probe pulse arrives at the sample before the pump pulse. Error bars reported
below correspond to $95\%$ confidence interval limits.

\begin{figure}
  \includegraphics[width=8cm,origin=c,angle=0]{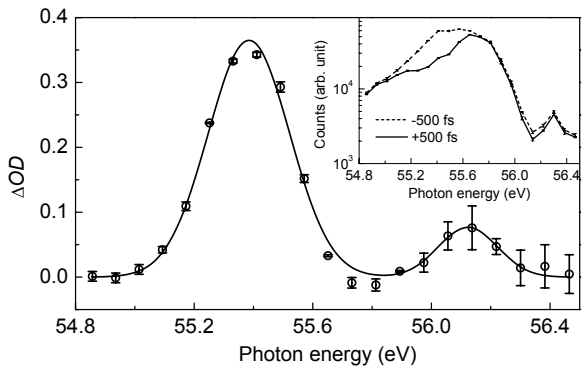}
  \caption[]{Transient absorption spectrum of Xe$^{+}$ acquired at a time delay of
           $+500$~fs, obtained from the average of 3 data sets. $\Delta OD$ denotes the
           change in optical density (absorbance). The
           $^2P_{3/2}\rightarrow\phantom{}^2D_{5/2}$  and
           $^2P_{1/2}\rightarrow\phantom{}^2D_{3/2}$ fine-structure transitions are located
           at $55.4$~eV and $56.1$~eV, respectively. The fit to a sum of two Gaussian
           curves is depicted by the solid line. The inset shows the harmonic spectra
           obtained at $-500$~fs and $+500$~fs pump-probe time delay.}
           \label{fig3}
\end{figure}

\begin{figure}
  \includegraphics[width=8cm,origin=c,angle=0]{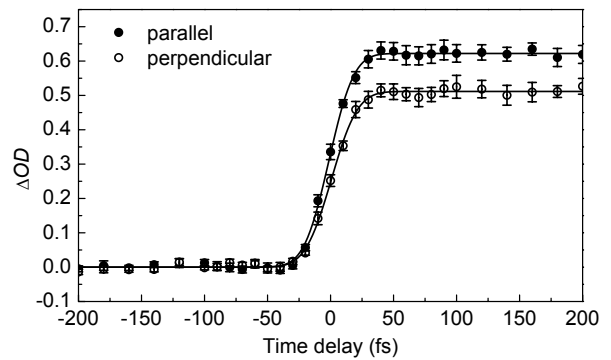}
  \caption[]{Time evolution of the $^2P_{3/2}\rightarrow\phantom{}^2D_{5/2}$
           transition for parallel and perpendicular relative polarizations between the
           optical pump and EUV probe pulses, obtained from the average of 8 time scans for
           each relative polarization. The polarization anisotropy observed at positive
           time delays implies that the $^{2}P_{3/2}$ state of Xe$^{+}$ produced by
           strong-field ionization is aligned.}
  \label{fig4}
\end{figure}

The transient absorption spectrum of Xe$^{+}$ acquired at a pump-probe time
delay of $+500$~fs and a parallel relative polarization between pump and probe
beams is shown in Fig.~\ref{fig3} \cite{dip}. The peaks at 55.4~eV and 56.1~eV
correspond to the $5p^{-1}_{3/2} (^2P_{3/2}) \rightarrow 4d^{-1}_{5/2}
(^2D_{5/2})$ and $5p^{-1}_{1/2} (^2P_{1/2}) \rightarrow 4d^{-1}_{3/2}
(^2D_{3/2}) $ transitions, respectively, in agreement with literature values
\cite{West01} ($nl^{-1}_{j}$ symbolizes a hole in the $nl$ orbital with angular
momentum $j$; the corresponding term symbol is given in parentheses). The ratio
of the areas of the two fine-structure absorption lines, defined as $R =
I^{\parallel}_{3/2 \rightarrow 5/2}/I^{\parallel}_{1/2 \rightarrow 3/2}$, is
found to be $R = 6.5 \pm 1.1$. The time-evolution of the Xe$^+$ $^2P_{3/2}$
state is followed by varying the pump-probe time delay while monitoring the
transient absorption signal at 55.4 eV. The resultant time traces obtained for
parallel and perpendicular relative polarizations between pump and probe beams
are shown in Fig.~\ref{fig4}.  Fitting the time traces to a convolution of a
step function with a Gaussian yields fwhm values of $37 \pm 1$ fs and $39 \pm 2$
fs for parallel and perpendicular relative polarizations, respectively. Note
that the temporal signal corresponds to a cross-correlation of the Xe$^{+}$
population growth with the EUV pulse \cite{pulsewidth}. From the
polarization-dependent absorption at positive time delays $\geq 50$~fs, the
polarization anisotropy, defined as $r = ({I_{3/2 \rightarrow 5/2}^{\parallel} -
I_{3/2 \rightarrow 5/2}^{\perp}})/({I_{3/2 \rightarrow 5/2}^{\parallel} + 2
I_{3/2 \rightarrow 5/2}^{\perp}})$, is found to be $r = 0.07 \pm 0.01$. The
observed anisotropy implies the existence of hole-orbital alignment in the
Xe$^{+}$ $^2P_{3/2}$ state produced by strong-field ionization. Since the hole
orbital is directed along the polarization axis of the pump beam (which also
defines the quantization axis), the hole population of the $m = \pm 1/2$
sublevels is expected to be greater than that of the $m = \pm 3/2$ sublevels, as
observed experimentally. To verify the reliability of the measured polarization
anisotropy, a separate set of measurements is performed for the
$^2P_{1/2}\rightarrow\phantom{}^2D_{3/2}$ transition, for which no anisotropy is
observed. This result agrees with the fact that alignment cannot exist in a
$^2P_{1/2}$ state \cite{Edmonds}.

The experimental results for the ratio of the fine-structure absorption $R$ and
the polarization anisotropy $r$ can be used to extract the complete quantum
state distribution. The description of the EUV-probe step follows
Ref.~\cite{SaDu06}. Let $\rho_{j,|m|}$ denote the probability of finding
Xe$^{+}$ with a hole in either the  $5p_{j,m}$ or the $5p_{j,-m}$ orbital.
Making the dipole approximation and employing standard angular momentum algebra
\cite{Edmonds}, we obtain
\begin{equation}
  \label{eq2}
  r = \frac{1}{10} \frac{\rho_{3/2,1/2} - \rho_{3/2,3/2}}{\rho_{3/2,1/2} +
  \rho_{3/2,3/2}}
\end{equation}
and
\begin{equation}
  \label{eq3}
  R = \frac{3 \rho_{3/2,1/2} + 2 \rho_{3/2,3/2}}{5 \rho_{1/2,1/2}} \xi,
\end{equation}
where $\xi = {\left| \langle ^2D_{5/2} \phantom{} \parallel d \parallel
\phantom{} ^2P_{3/2} \rangle\right|^2} / {\left|\langle ^2D_{3/2} \phantom{}
\parallel d \parallel \phantom{} ^2P_{1/2} \rangle\right|^2}$ and $d$ is the
electric dipole operator. A multiconfiguration Dirac-Fock calculation performed
with the program package {\sc GRASP2} \cite{PaFr96} gives $\xi=1.6$. From
Eqs.~(\ref{eq2}) and (\ref{eq3}), and the experimental values for $R$ and $r$,
the complete quantum state distribution of Xe$^+$ generated by strong-field
ionization can be extracted. The results are summarized in
Table~\ref{tab:table1}. We note that spectroscopic probing of the photoion by
transient absorption allows direct retrieval of its quantum state distribution,
which cannot be obtained from other experimental approaches to the study of
strong-field ionization, such as energy- and angular-resolved photoelectron
spectroscopy \cite{DiMauro96}.

Tunnel ionization calculations with the inclusion of spin-orbit coupling are
performed to model the experimental results described above. To calculate the
production rates for the various ionization channels, the effective one-electron
model described in Ref.~\cite{SaDu06} is employed. The model treats strong-field
ionization within the tunneling picture and includes the effect of spin-orbit
interaction. The approach makes use of a flexible finite-element basis set and
determines ionization rates in this square-integrable basis using a complex
absorbing potential \cite{RiMe93}.  Parameters that are used in the calculations
are 0.13 mJ for the pulse energy, 49 fs fwhm for the pulse duration, and 45
$\mu$m for the cylindrically-symmetric beam waist. The $\vert j,m \rangle$
populations are calculated after the laser pulse by numerical integration of
rate equations \cite{SaDu06} and are normalized such that $\rho_{3/2,1/2} +
\rho_{1/2,1/2} + \rho_{3/2,3/2} = 1$. The calculated fractional populations on
the laser axis are shown in Table~\ref{tab:table1}. These calculations also
allow us to verify that effects due to spatial averaging by the probe beam are
smaller than the experimental errors for the $\rho_{j,|m|}$ distribution.

Within the given uncertainties, good agreement for the
$\rho_{3/2,1/2}:\rho_{1/2,1/2}$ ratio is obtained between experiment and theory.
This suggests that spin-orbit interaction is adequately treated in the tunnel
ionization model employed in the calculations. Moreover, the calculations reveal
that there is very little mixing between the $5p_{3/2}$ and $5p_{1/2}$ valence
orbitals even at the saturation intensity for Xe$^{+}$ production. This implies
that the laser field is not sufficiently strong to uncouple the spin-orbit
interaction.

\begin{table}
  \caption{\label{tab:table1}Comparison of the complete $\vert j, m \rangle$
  quantum state distribution obtained from experiment and theory for  Xe$^+$
  generated via strong-field ionization.}
  \begin{ruledtabular}
    \begin{tabular}{ccc}
      &\multicolumn{2}{c}{Population distribution $\rho_{j,\vert m \vert} (\%)$ }\\
      $\vert j, m \rangle$ & Experimental & Theoretical\\ \hline
      $\vert \frac{3}{2}, \pm \frac{1}{2} \rangle$ & $75 \pm 6$ & 83\\
      $\vert \frac{1}{2}, \pm \frac{1}{2} \rangle$ & $12 \pm 3$ & 14\\
      $\vert \frac{3}{2}, \pm \frac{3}{2} \rangle$ & $13 \pm 6$ & 3\\
    \end{tabular}
  \end{ruledtabular}
\end{table}

Furthermore, the tunneling calculation predicts that $\rho_{3/2,3/2}\ll
\rho_{3/2,1/2}$, which supports the experimental observation that the Xe$^{+}$
$^2P_{3/2}$  state produced by strong-field ionization is aligned. However, it
is notable that the measured $\rho_{3/2,3/2}:\rho_{3/2,1/2}$ ratio of $0.17 \pm
0.09$ is significantly larger than that predicted by the calculation, which
gives a value of 0.04 for this ratio. The discrepancy between experiment and
theory can be attributed to the partial breakdown of the adiabatic
(quasi-static) approximation employed in the tunnel ionization model.

The adiabatic approximation requires that the electrons within the atomic
potential and those undergoing tunnel ionization respond instantaneously to the
laser field. This approximation is analogous to the Born-Oppenheimer
approximation frequently invoked in the study of molecular dynamics, whereby
electrons are assumed to respond instantaneously to the electric field exerted
by the nuclei. The ratio of the tunneling time to the period of the laser field
is given by the Keldysh adiabaticity parameter $\gamma$ \cite{Keldysh65},
implying that the adiabatic approximation is valid only when $\gamma \ll 1$.
Given that the experimental conditions yield  $\gamma = \sqrt{I_p/2 U_p} \sim
1.1$  ($I_p$ is the atomic ionization potential and $U_p$ is the ponderomotive
potential), such a quasi-static approximation is no longer wholly valid, i.e.,
ionization becomes nonadiabatic with respect to the laser field. Under such
circumstances, a nonperturbative multiphoton Floquet treatment of strong-field
ionization becomes necessary \cite{Floquet}. Indeed, previous investigations of
Xe strong-field ionization revealed discrete peaks atop above-threshold
ionization features in the photoelectron spectra, suggesting that ionization
occurs mainly via a multiphoton pathway \cite{Mevel93,Witzel00}; the discrete
peaks originate from resonance-enhanced multiphoton ionization that occurs when
the field-dressed ground state crosses the ac Stark-shifted Rydberg states
\cite{Freeman87}. On the sub-cycle timescale, nonadiabatic transitions can occur
between the field-dressed ground state and the Rydberg levels while varying the
instantaneous phase of the laser field, akin to transitions between adiabatic
potential energy surfaces in the non-Born-Oppenheimer regime of molecular
dynamics. Note that nonadiabatic electron dynamics have previously been observed
in molecular strong-field ionization, in which the spatial delocalization of
electrons leads to the breakdown of the adiabatic approximation \cite{Stolow01}.

In the present work, the importance of nonadiabatic effects in atomic
strong-field ionization is substantiated by the results of two spin-orbit-free
calculations---a tunneling calculation and a multiphoton Floquet-type
calculation \cite{BuSa06}---performed at an intensity of $8\times
10^{13}$~W/cm$^2$. The Floquet calculation predicts that the ratio of the $m_l =
\pm 1 : m_l = 0$ ionization rates is $\sim 2\times$ larger than that predicted
by the tunnel ionization model. Therefore it is reasonable that the
experimentally-measured $\rho_{3/2,3/2}:\rho_{3/2,1/2}$ ratio is larger than
that predicted by the tunneling calculation with spin-orbit coupling. A valid
theoretical comparison with the experimental result would require a Floquet
calculation that incorporates spin-orbit coupling.

Finally, we note that while recent advances in high-order harmonic generation
have already resulted in the extension of ultrafast spectroscopy into the EUV
domain \cite{Leone01,KaMu01,Heinzmann01,Krausz02}, most of the experiments
reported to date have focused on photoelectron spectroscopy due to the set of
discrete harmonics that presents itself as an attractive photoionization source.
By probing transitions from the core level to unoccupied valence levels,
picosecond time-resolved x-ray absorption near-edge spectroscopy based on
laser-produced plasma \cite{Wilson96} and synchrotron \cite{Bressler04,Chen05}
sources have been shown to be highly sensitive to element-specific electronic
structure changes accompanying photophysical and photochemical transformation.
The work here demonstrates the feasibility of performing transient absorption
spectroscopy with sub-50-fs time resolution using a laser-based, high-order
harmonic source for core-level probing. Work on extending this technique to the
study of ultrafast molecular dynamics is currently in progress.

We thank T. Pfeifer for useful discussions. The assistance from A. Paul and
Profs. M. M. Murnane and H. C. Kapteyn with implementing the HHG setup is
gratefully acknowledged. M.K. was supported by a Miller Fellowship. This work
was supported by the NSF ERC for EUV Science and Technology (EEC-0310717) and
the LDRD program at LBNL, with additional equipment and support from DOE
(DE-AC02-05CH11231). R.S. was supported by DOE (DE-AC02-06CH11357). C.B. was
funded by a Feodor Lynen Research Fellowship from the Alexander von Humboldt
Foundation.


\begin{thebibliography}{99}
\bibitem{pressure}
From the estimated electron rms velocity (1 $\mu$m/ps, corresponding to an
electron temperature of $\sim 10$~eV for the plasma) and typical electron-ion
scattering cross-sections ($\sim 10^{-16}$ cm$^{2}$), the time interval between
each electron-ion collision event for the given sample density is $\sim 100$~ps.
Since this time is much longer than the maximum pump-probe time delay employed
in the experiments, the observed dynamics are those of single ions, and the
measured alignment distribution is not affected by electron-ion or ion-ion
collisions.
\bibitem{dip}
The minimum at 56.1~eV in the harmonic spectrum collected at $-500$~fs time
delay exists in the weak continuum underlying the discrete harmonic peaks (note
the use of a log scale); this could result from quantum interferences and
macroscopic phasematching effects in HHG.
\bibitem{pulsewidth}
From the measured rise times and the computed time-evolution of $\rho_{3/2,1/2}$
and $\rho_{3/2,3/2}$, the EUV probe pulse duration is estimated to be $\sim 30$
fs.
\end{thebibliography}
\end{document}